\documentclass{ws-ijmpa}
\usepackage[super,compress]{cite}
\usepackage{cancel}
\usepackage{graphicx}
\usepackage{hyperref}
\usepackage{extarrows}
\usepackage{amsmath}
\usepackage{amsfonts}
\usepackage{amssymb}
\usepackage{float}
\begin{document}
	\markboth{Pritam Das, Mrinal Kumar Das}{Phenomenology of $keV$ sterile neutrino in minimal extended seesaw}
	
	%
	\catchline{}{}{}{}{}
	%
	\title{Phenomenology of $keV$ sterile neutrino in minimal extended seesaw}
	\author{Pritam Das }
	\address{ Department of Physics, Tezpur University, Assam - 784\,028, India\\pritam@tezu.ernet.in}
	\author{Mrinal Kumar Das}
	\address{%
		Department of Physics, Tezpur University, Assam - 784\,028, India \\	mkdas@tezu.ernet.in%
	}%

\maketitle
\begin{history}
\end{history}

	\begin{abstract}
We explore the possibility of a single generation of $keV$ scale sterile neutrino ($m_S$) as a dark matter candidate within the minimal extended seesaw (MES) framework and it's influence in neutrinoless double beta decay ($0\nu\beta\beta$) study. Three hierarchical right-handed neutrinos were considered to explain neutrino mass. We also address baryogenesis via the mechanism of thermal leptogenesis considering the decay of the lightest RH neutrino to a lepton and Higgs doublet. A generic model based on $A_4\times Z_4\times Z_3$ flavor symmetry is constructed to explain both normal and inverted hierarchy mass pattern of neutrinos. Significant results on effective neutrino masses are observed in presence of sterile mass ($m_S$) and active-sterile mixing ($\theta_{S}$) in $0\nu\beta\beta$. Results from $0\nu\beta\beta$ give stringent upper bounds on the active-sterile mixing matrix element. To establish sterile neutrino as  dark matter within this model, we checked decay width and relic abundance of the sterile neutrino, which restricted sterile mass ($m_S$) within some definite bounds. Constrained regions on the CP-phases and Yukawa couplings are obtained from $0\nu\beta\beta$ and baryogenesis results.  Co-relations among these observable are also established and discussed within this framework.
\end{abstract}
\ccode{PACS numbers:}
\section{INTRODUCTION}
The discovery of neutrino mass and the Higgs Boson have brought glory to the field of particle physics as well as to astrophysics and cosmology. Experimental results in the field of neutrinos \cite{Abe:2016nxk, boger2000sudbury,evans2013minos,Abe:2011sj,Ahn:2012nd, Abe:2011fz,An:2012eh} not only verify the theoretical predictions but also open up a new portal to bring physics to the next level\footnote{Recent global fit results with $3\sigma$ bound and best-fit values of the observed neutrino parameters are given in the tabular form in table \ref{tab:d1}.}. In spite of the glorious successes, many unsettled phenomenons and queries are still around us. Exact nature and absolute mass scale of the neutrinos, matter-antimatter asymmetry of the Universe, presence of extra flavor of neutrinos, Dark Matter, {\it etc.} are among them.\\ Recent results from several cosmological observations \cite{Athanassopoulos:1997pv, Athanassopoulos:1996jb, Aguilar:2001ty} as well as reactor data \cite{Abdurashitov:2005tb, giunti2011statistical, giunti2012update} reveal the fact that heavy flavor of neutrinos do exist in the Universe, and they are known as {\it sterile neutrino}. Sterile neutrinos are neutral right-handed (RH) fermions, and they are singlets under the SM gauge group. Unlike the active neutrinos, they are infertile, {\it i.e.,} they do not change flavor; however, they mix up with the active neutrinos. For better understanding of sterile neutrino nature and interactions, one may refer to the well established works in\cite{Dodelson:1993je,Abazajian:2012ys, Abazajian:2017tcc, Benes:2005hn, Barry:2011wb}. Despite the fact that, the exact mass scale or numbers of sterile neutrino generations are still unknown, their presence may have a significant contribution to the new physics. The presence of sterile neutrino is strongly motivated and highly influences the current reactor neutrino anomalies. Sterile neutrino with different mass ranges play crucial role in astrophysics \cite{Petraki:2007gq}, cosmology \cite{Abazajian:2012ys,Abazajian:2017tcc}, collider physics \cite{Abada:2017jjx,Atre:2009rg,Deppisch:2015qwa}, etc. Similar kind of studies were carried out in other context such as LRSM \cite{Borgohain:2019pya,Barry:2014ika}, extra dimensions \cite{Rodejohann:2014eka, Dev:2012bd}, in presence of exotic charged currents \cite{Ludl:2016ane} or in relation with $keV$ neutrino dark matter \cite{deVega:2011xh}. \\
Absolute neutrino mass is yet another unknown to the physics community as oscillation experiments are only sensitive to the mass squared difference ($\Delta m_{ij}^2$) and leptonic mixing angles ($\theta_{ij}$, with $i,j=1,2,3$). Apart from the oscillation studies, the kinematic study of reactions involving neutrino ($\nu$) and anti-neutrino ($\overline{\nu}$) can give us information about absolute mass. Considering Majorana nature of particles, Wendell Furry \cite{1939PhRv...56.1184F} studied a kinetic process similar to "double-beta disintegration" without neutrino emission, popularly known as {\it neutrino-less double beta decay} $(0\nu\beta\beta)$ \cite{DellOro:2016tmg}. In simple word this can be expressed as, $$(A,Z)\rightarrow(A,Z+2)+2e^-.$$
From $0\nu\beta\beta$ integration, if Majorana nature of the neutrino is verified, one can give conclusive remark on absolute neutrino mass.
The  $(0\nu\beta\beta)$ process explicitly violets the lepton number by creating a pair of electron. Discovery of lepton number violation (LNV) process supported by existing theoretical picture and the $0\nu\beta\beta$ scenario allows leptons to take part in the process of matter-antimatter asymmetry of our Universe. Thus the observation of such a process is crucial for demonstrating baryogenesis idea \cite{Cline:2006ts} via lepton number violation. Many works on  $(0\nu\beta\beta)$ have been done considering the SM neutrinos \cite{Bilenky:2004wn,Bilenky:2012qi,Agostini:2018tnm,Borgohain:2017akh}. Nevertheless, it is now clear that the addition of a new scalar fermion and study its interactions within the SM particles can lead us to a broad range of new physics phenomenology \cite{Barry:2011wb,Abada:2018qok}.\\
Shreds of evidence from various sources \cite{pecontal2009review, clowe2006direct, bennett2013nine, Ade:2013zuv}, it now confirmed the presence of dark matter (DM) into the picture. To understand DM and their mysterious behaviour, we have to understand what it composed of and how they interact with known particles.
Among the choices of being a dark matter candidate, dense baryonic and non-baryonic matters were first proposed, and they are largely disfavoured \cite{clowe2006direct, Yoo:2003fr, Pani:2014rca}. Modifications to the laws of gravitational~\cite{Milgrom:1983ca} were also not so impressive to explain DM. Since, no SM particle can be a dark matter candidate \cite{Kraus:2004zw, Lobashev:2001uu, Kolb:1990vq}, so addition of a new particle to the elementary particle list was the only reasonable choice. 
In the past years, several particles were proposed as DM candidates in BSM and WIMPs (weakly interacting massive particles) are the most attractive candidates for dark matter at current scenario. WIMPs do not create problems in structure formation like the SM neutrinos does, due to non-relativistic velocity and higher masses. They take different forms under different scenarios like neutralinos under SUSY \cite{Jungman:1995df,Gelmini:2006pw}, Kaluza-Klein bosons as predicted by models based on extra spatial dimensions \cite{Servant:2002aq,Bonnevier:2011km} and minimal extension of the SM scalar sector consider inert doublet scalar as WIMP DM \cite{LopezHonorez:2006gr,Khan:2017xyh, Das:2019ntw}.\\
Apart from DM and absolute neutrino mass, the overabundance of baryonic matter over the anti-baryonic matter is also discussed in this work. The baryon asymmetry of the Universe (BAU) ($Y_B\equiv\frac{n_B-n_{\overline{B}}}{s}\sim (8.7\pm0.06)\times 10^{-11}$)\footnote{$n_B$ and $n_{\overline{B}}$ are the baryon and anti-baryon number density respectively. $s$ in the denominator is the entropy of the current Universe.}\cite{Aghanim:2018eyx} is well explained by baryogenesis \cite{Fukugita:1986hr,Strumia:2006qk}. Rich literatures available in \cite{Davidson:2008bu,DiBari:2012fz,Nardi:2006fx, Buchmuller:2004tu,Frossard:2012pc, Borah:2015vra, Kalita:2014vxa, Borah:2013bza, Borgohain:2017akh} discussing baryogenesis via the mechanism of thermal leptogenesis. We considered thermal leptogenesis, where the heavy RH neutrinos are hierarchical ($M_{\nu_{R1}}<M_{\nu_{R_{2,3}}}$). As per our preferred choice of  mass for lightest RH neutrinos \cite{Davidson:2008bu}, we are restricted our study to a single flavor leptogenesis. \\
Motivated by these studies, we are considering a sterile neutrino flavor with a mass around $keV$ range in minimal extended seesaw (MES) \cite{Barry:2011wb,Zhang:2011vh,Rodejohann:2014eka, Das:2018qyt}, where an additional fermion singlet (sterile neutrino) is added along with three RH neutrinos. The beauty of MES framework is that it can accommodate sterile neutrino mass ranging from eV to $keV$. Sterile neutrino with eV as well as $keV$ could be probed in future KATRIN experiment \cite{Osipowicz:2001sq,Mertens:2014nha}. Moreover, $keV$ sterile neutrino has a potential to affect electron energy spectrum in tritium $\beta$-decays \cite{Shrock:1980vy}. Typically, sterile neutrinos with mass (0.4-50) $keV$ \cite{Boyarsky:2009ix} are considered as WIMP particles since they are relatively slow and much heavier compared to the active neutrinos. In fact for successfully observe $0\nu \beta\beta$ the upper bound for sterile neutrino mass should be 18.5 $keV$ \cite{Abada:2018qok, Benso:2019jog}. Back in the 90s, Dodelson
and Widrow \cite{Dodelson:1993je} proposed $keV$ sterile neutrino as dark matter candidate produced via oscillation and collision from active neutrinos. Recently from various sources like, in the stacked spectrum of galaxy clusters \cite{Bulbul_2014}, individual spectra of nearby galaxy clusters \cite{Bulbul_2014,PhysRevLett.113.251301}, Andromeda Galaxy \cite{PhysRevLett.113.251301}, and in the Galactic Center region \cite{Boyarsky:2014ska,Riemer-Sorensen:2014yda} an unidentified line was reported. The position of the line is $E = 3.55$ $ keV$ with an uncertainty in position $\simeq\pm 0.05$ $keV$. If the line is interpreted as originating from a two-body decay of a DM particle, then the particle has its mass at about $m_S\simeq 7.1$ $keV$ and the lifetime $\tau_{DM} \simeq 10^{27.8\pm 0.3}$ sec \cite{PhysRevLett.113.251301}. 
Hence, choosing a mass range for the $keV$ regime sterile neutrino within (1-18.5) $keV$ and explore new possibilities to explain $0\nu\beta\beta$ within laboratory constraints along with DM signature is quite a good choice. Along with the sterile study, we also try to verify baryogenesis produced via the mechanism of thermal leptogenesis within our model and finally, we try to co-relate all these observable under the same framework. 
\\
This work is organized as follows: model building with $A_4$ flavor symmetry along with $Z_4\times Z_3$ discrete symmetry is discussed in section \ref{ch2} for both normal (\ref{cnh}) and inverted (\ref{cih}) mass pattern. Numerical analysis is carried out in section \ref{num} and separate sub-section for $0\nu\beta\beta$ (\ref{ndbd}), dark matter (\ref{sdm}) and baryogenesis via thermal leptogenesis (\ref{lep}) are carried out under the same numerical analysis section in a respective manner. Results of our study are discussed in section \ref{result} and finally we conclude our work in section \ref{conc}. 
			\begin{table}[ph]
			\tbl{Recent experiments results for active neutrinos parameters with best-fit and the latest global fit $3\sigma$ range \cite{Capozzi:2016rtj}.}
			{\begin{tabular}{@{}ccc@{}} \toprule
	Parameters&NH (Best fit) & IH (Best fit)\\
			\hline
			$\Delta m^{2}_{21}[10^{-5}eV^2]$&6.93-7.97(7.73)&6.93-7.97(7.73)\\
			$\Delta m^{2}_{31}[10^{-3}eV^2]$&2.37-2.63(2.50)&2.33-2.60(2.46)\\
			$sin^{2}\theta_{12}/10^{-1}$&2.50-3.54(2.97)&2.50-3.54(2.97)\\
			$sin^{2}\theta_{13}/10^{-2}$&1.85-2.46(2.14)&1.86-2.48(2.18)\\ 
			$sin^{2}\theta_{23}/10^{-1}$&3.79-6.16(4.37)&3.83-6.37(5.69)\\
			$\delta_{13}/\pi$&0-2(1.35)&0-2(1.32)\\ \hline
		\end{tabular}\label{tab:d1}}
\end{table}
\section{The Model}\label{ch2}
\subsection{Normal Hierarchy:}\label{cnh}
In neutrino model building phenomenology, symmetries have been playing an important role in describing various phenomenology. Interestingly, discrete symmetries like $A_4$ with $Z_n$ ($n\ge2$ is integer) are much popular in recent literature in explaining neutrino mass  \cite{Altarelli:2005yp,Chun:1995bb,Barry:2011wb,Babu:2009fd}. The discrete flavor symmetry $A_4$ being the symmetry group of rotation with a tetrahedron invariant with 4 irreducible representation denoted by $\bf{1},\bf{1^{\prime}},\bf{1^{\prime\prime}} $ and $\bf{3}$. The left-handed (LH) lepton doublet $l$ to transform as $A_4$ triplet whereas the right-handed (RH) charged leptons ($e^c,\mu^c,\tau^c$) transform as 1,$1^{\prime\prime}$ and $1^{\prime}$ respectively.  Apart from the type-I seesaw particle content, few extra flavons are added to construct the model.
Two triplets $\zeta, \varphi$, two singlets $\xi$ and $\xi^{\prime}$ are added to produce broken flavor symmetry. Besides the SM Higgs $H_1$, we also introduce an additional Higgs doublets ($H_2$) \cite{Felipe:2013vwa,Nath:2016mts}, to make the model work. Non-desirable interactions were restricted using extra $Z_4$ and $Z_3$ charges to the fields. To accommodate sterile neutrino into the framework, we add a chiral gauge singlet $S$, which interacts with the RH neutrino $\nu_{R1}$ via $A_4$ singlet ($1^{\prime}$) flavon $\chi$ to give rise to sterile mixing matrix. We used dimension-5 operators \cite{weinberg} for Dirac neutrino and charged lepton mass generation. One may notice, terms like $\frac{1}{\Lambda} \overline{S^c}S\varphi\varphi$ may ruin the current MES scenario, by giving rise to unexpectedly higher mass term for the sterile neutrino \cite{Zhang:2011vh}. Those terms are excluded by the $Z_3$ symmetry.

As per the MES structure, the newly added singlet field $S$ does not interact with the active neutrinos, and they can be explained with the Abelian symmetries. For example, by introducing additional $U(1)^{\prime}$ charge under which the SM particles and RH neutrinos are to be neutral. The singlet $S$ on the other hand carries a $U(1)^{\prime}$ charge $Y^{\prime}$ and we further introduced a SM singlet $\chi$ with hypercharge $-Y^{\prime}$. Hence those coupling of $S$ with the active neutrinos are still forbidden by the $U(1)^{\prime}$ symmetry at the renormalizable level \cite{Barry:2011wb, Chun:1995js, Heeck:2012bz}. 
The particle content with  $A_4 \times Z_4\times Z_3$ charge assignment under NH are shown in the table \ref{tab1}.\\

	\begin{table}[h]
	\tbl{Particle content and their charge assignments under SU(2), $A_4$ and $Z_4\times Z_3$ groups for NH mode.}
	{\begin{tabular}{@{}cccccccccccccccccc@{}} \toprule
			Particles&  $l$ &  $e_{R}$&  $\mu_{R}$&  $\tau_{R}$&  $H_1$&  $H_2$&  $\zeta$&  $\varphi$&  $\xi$&  $\xi^{\prime}$&  $\nu_{R1}$&  $\nu_{R2}$&  $\nu_{R3}$&  $S$&  $\chi$\\
			\hline
			SU(2)&2&1&1&1&2&2&1&1&1&1&1&1&1&1&1\\
			$A_4$&3&1&$1^{\prime\prime}$&$1^{\prime}$&1&1&3&3&1&$1^{\prime}$&1&$1^{\prime}$&1&$1^{\prime\prime}$&$1^{\prime}$\\
			$Z_4$&1&1&1&1&1&i&1&i&1&-1&1&-i&-1&i&-i\\
			$Z_3$&1&1&1&1&1&1&1&1&1&1&1&1&1&$\omega^2$&$\omega$\\
			\hline
		\end{tabular}\label{tab1}}
\end{table}
In lepton sector, the leading order invariant Yukawa Lagrangian is given by,
\begin{equation}\label{lag}
\begin{split}
\mathcal{L} =& \frac{y_{e}}{\Lambda}(\overline{l} H_1 \zeta)_{1}e_{R}+\frac{y_{\mu}}{\Lambda}(\overline{l} H_1 \zeta)_{1^{\prime}}\mu_{R}+\frac{y_{\tau}}{\Lambda}(\overline{l} H_1 \zeta)_{1^{\prime\prime}}\tau_{R} \\
&+ \frac{y_{2}}{\Lambda}(\overline{l}\tilde{H_1}\zeta)_{1}\nu_{R1}+\frac{y_{2}}{\Lambda}(\overline{l}\tilde{H_1}\varphi)_{1^{\prime\prime}}\nu_{R2}+\frac{y_{3}}{\Lambda}(\overline{l}\tilde{H_2}\varphi)_{1}\nu_{R3}\\
& +\frac{1}{2}\lambda_{1}\xi\overline{\nu^{c}_{R1}}\nu_{R1}+\frac{1}{2}\lambda_{2}\xi^{\prime}\overline{\nu^{c}_{R2}}\nu_{R2}+\frac{1}{2}\lambda_{3}\xi\overline{\nu^{c}_{R3}}\nu_{R3}\\
&+ \frac{1}{2}\rho\chi\overline{S^{c}}\nu_{R1} .\\
\end{split}
\end{equation}
In this Lagrangian, various Yukawa couplings are represented by $y_{\alpha,i}$, $\lambda_{i}$ (for $\alpha=e,\mu,\tau$ and $i=1,2,3$) and $\rho$ for respective interactions. Higgs doublets are transformed as $\tilde{H} = i\tau_{2}H^*$ ($\tau_{2}$ is the second Pauli's spin matrix) to keep the Lagrangian gauge invariant and $\Lambda$ is the cut-off scale of the theory, which is around the GUT scale. 
The scalar flavons involved in the Lagrangian acquire VEV along  $ \langle \zeta \rangle=(v,0,0), \langle\varphi\rangle=(v,v,v), 
\langle\xi\rangle=\langle\xi^{\prime}\rangle=v$ and $ \langle\chi\rangle=v_{\chi}$ by breaking the flavor symmetry, while $\langle H_i\rangle(i=1,2)$ get VEV ($v_i$) by breaking EWSB at electro-weak scale
Following the $A_4$ product rules and using the above mentioned VEV alignment\footnote{The triplet VEV alignment of the scalars are the solution of the respective scalars at their minimal potential \cite{Das:2018qyt}}, the Dirac neutrino mass\footnote{ $M_D^{\prime}$ is the unmodified Dirac neutrino mass matrix which is failed to generate $\theta_{13}\neq 0$. The modified $M_D$ is represented in equation \eqref{md} }, Majorana neutrino mass and the sterile mass matrices are given by,
\begin{equation}
M^{\prime}_{D}=
\begin{pmatrix}
D_1&D_1&D_2\\
0&D_1&D_2\\
0&D_1&D_2\\
\end{pmatrix},\ M_{R}=\begin{pmatrix}
R_1&0&0\\
0&R_2&0\\
0&0&R_3\\
\end{pmatrix}, \
M_{S}= \begin{pmatrix}
G&0&0\\
\end{pmatrix}.
\label{emd}
\end{equation}
where, $D_1=\frac{\langle H_1\rangle v}{\Lambda}y_{2} $ and $D_2=\frac{\langle H_2\rangle v}{\Lambda}y_{3}$\footnote{We have assumed the VEV of the Higgs doublets to be identical for simplicity.}. Other elements are defined as $R_1=\lambda_{1}v, R_2=\lambda_{2}v$, $R_3=\lambda_{3}v$ and $G=\rho v_\chi$.  In order to achieve sterile mass in the $keV$ range, we have considered VEV for the $\chi$ flavon lie aroud TeV scale. A rough estimate of the mass scales of parameters are given as , $\Lambda\simeq 10^{14}$ GeV, $v\simeq 10^{13}$ GeV and $v_{\chi}\simeq 10$ TeV. \\
We have used similar approaches from our previous work \cite{Das:2018qyt} to break the trivial $\mu-\tau$ symmetry in the light neutrino mass matrix. We introduced two new $SU(2)$ singlet flavon fields ($\zeta^{\prime}$ and $\varphi^{\prime}$)  which results the $M_P$ matrix \eqref{pmatrix} when they couple with the respective RH neutrinos. The active mass matrix gets modify by adding the matrix \eqref{pmatrix} to the Dirac neutrino mass matrix. This new $M_P$ matrix played a significant role in producing non-zero reactor mixing angle and has potential influence in choosing the octant for $\theta_{23}$ \cite{Das:2018qyt}. The Lagrangian that generate the matrix \eqref{pmatrix} can be written as, 
\begin{equation}
\mathcal{L}_{\mathcal{M_P}} =\frac{y_{1}}{\Lambda}(\overline{l}\tilde{H_1}\zeta^{\prime})_{1}\nu_{R1}+\frac{y_{1}}{\Lambda}(\overline{l}\tilde{H_1}\varphi^{\prime})_{1^{\prime\prime}}\nu_{R2}+\frac{y_{1}}{\Lambda}(\overline{l}\tilde{H_2}\varphi^{\prime})_{1}\nu_{R3}.
\end{equation}
New $SU(2)$ singlet flavon fields ($\zeta^{\prime}$ and $\varphi^{\prime}$) are considered and supposed to take $A_4\times Z_4\times Z_3$ charges as same as $\zeta$ and $\varphi$ respectively. After breaking flavor symmetry they acquire VEV along $\langle\zeta^{\prime}\rangle=(v_p,0,0)$ and $\langle\varphi^{\prime}\rangle=(0,v_p,0)$ directions, giving rise to the $M_P$ matrix as,
\begin{equation}\label{pmatrix}
M_{P}=
\begin{pmatrix}
0&0&P\\
0&P&0\\
P&0&0\\
\end{pmatrix},
\end{equation}
with, $P=\frac{\langle H_{i}\rangle v}{\Lambda}y_{1}$ ($i=$1 or 2).
Scale of these VEV ($v_p$) in comparison to earlier flavon's VEV ($v$) are differ by an order of magnitude ($v>v_p$). Involvement of these new flavons are restricted in the leading order Lagrangian charge lepton mass matrix. Hence, the Dirac neutrino mass matrix, $M_D$ from eq. \eqref{emd} will take new structure as,
\begin{equation} \label{md}
M_D=M^{\prime}_D+M_P=
\begin{pmatrix}
D_1&D_1&D_2+P\\
0&D_1+P&D_2\\
P&D_1&D_2\\
\end{pmatrix}.
\end{equation}
\subsection{Inverted Hierarchy} \label{cih}
	\begin{table}[h]
	\tbl{Particle content and their charge assignments under SU(2), $A_4$ and $Z_4\times Z_3$ groups for IH.}
	{\begin{tabular}{@{}ccccccccccccccccccc@{}} \toprule
			Particles&  $l$ &  $e_{R}$&  $\mu_{R}$&  $\tau_{R}$&  $H_1$&  $H_2$&  $\zeta$&  $\varphi$& $\varphi^{\prime\prime} $&$\xi$&  $\xi^{\prime}$&  $\nu_{R1}$&  $\nu_{R2}$&  $\nu_{R3}$&  $S$&  $\chi$\\
			\hline
			SU(2)&2&1&1&1&2&2&1&1&1&1&1&1&1&1&1&1\\
			$A_4$&3&1&$1^{\prime\prime}$&$1^{\prime}$&1&1&3&3&1&$1^{\prime}$&1&1&$1^{\prime}$&1&$1^{\prime\prime}$&$1^{\prime}$\\
			$Z_4$&1&1&1&1&1&i&1&1&i&1&-1&1&-i&-1&i&-i\\
			$Z_3$&1&1&1&1&$\omega$&1&1&1&1&$\omega^2$&$\omega^2$&$\omega^2$&$\omega$&$\omega^2$&$\omega^2$&$\omega$\\
			\hline
		\end{tabular}\label{tab2}}
\end{table}
Within MES, the situation is not that simple for IH mode \cite{Das:2018qyt, Zhang:2011vh}. A slight change in VEV arrangement is required in IH mode in order to give correct observed phenomenology \cite{Zhang:2011vh}. A new triplet flavon $\varphi^{\prime\prime}$ with VEV alignment along $\langle\varphi^{\prime\prime}\rangle \sim (2v,-v,-v)$ is introduced, which modifies the Dirac neutrino mass matrix. Particles and charges under symmetry groups ($SU(2)\times A_4\times Z_4 \times Z_2$ ) are shown in table \ref{tab2}. The modified Yukawa Lagrangian for the $M_D$ is given by,
\begin{equation}
\mathcal{L}_{\mathcal{M_D}}= \frac{y_{2}}{\Lambda}(\overline{l}\tilde{H_1}\zeta)_{1}\nu_{R1}+\frac{y_{2}}{\Lambda}(\overline{l}\tilde{H_1}\varphi^{\prime\prime})_{1^{\prime\prime}}\nu_{R2}+\frac{y_{3}}{\Lambda}(\overline{l}\tilde{H_1}\varphi)_{1}\nu_{R3}.
\end{equation}
Except the Dirac Lagrangian, other Lagrangian will retain the same form as per the equation \ref{lag}. The Dirac neutrino mass matrix takes new structure as, 
\begin{equation}
M^{\prime}_{D}=
\begin{pmatrix}
D_1&-D_1&D_2\\
0&-D_1&D_2\\
0&2D_1&D_2\\
\end{pmatrix}.
\end{equation}
Similar to the NH case, this Dirac neutrino mass matrix also get modified by adding the $M_P$ matrix.  We have shown the complete matrix structure for both the mass ordering in the table \ref{tab:nh} \\
\section{Numerical Analysis}\label{num}
Following the minimal extended seesaw (MES) framework we set up active and sterile neutrino mass matrices. 
In MES scenario three extra right-handed neutrinos and one additional gauge singlet chiral field $S$ is introduced along with the SM particles. The MES Lagrangian for neutrino mass terms is given by,
\begin{equation}
-\mathcal{L}_{\mathcal{M}}= \overline{\nu_{L}}M_{D}\nu_{R}+\frac{1}{2}\overline{\nu^{c}_{R}}M_{R}\nu_{R}+\overline{S^c}M_{S}\nu_{R}+h.c. ,
\end{equation}  
Here, $M_D$ and $M_R$ are $3\times3$ Dirac and Majorana neutrino mass matrices respectively with $M_S$ being a $1\times3$ matrix.
A detailed discussion on MES has already been carried out in previous works \cite{Das:2018qyt, Barry:2011wb, Zhang:2011vh}. The active neutrino mass matrix within MES framework is given by,
\begin{equation}\label{amass}
m_{\nu}\simeq M_{D}M_{R}^{-1}M_{S}^T(M_{S}M_{R}^{-1}M_{S}^{T})^{-1}M_{S}(M_{R}^{-1})^{T}M_{D}^{T}-M_{D}M_{R}^{-1}M_{D}^{T}, 
\end{equation} and the $keV$ scaled sterile neutrino mass as,
\begin{equation}\label{smass}
m_{s}\simeq -M_{S}M_{R}^{-1}M_{S}^{T}.
\end{equation}
With these \eqref{amass} and\eqref{smass} equations, we established the active and sterile mass structures for both the NH as well as IH.\\
We diagonalize the active neutrino mass matrix using the popular $U_{PMNS}$ matrix \cite{Giganti:2017fhf}.
The diagonalize neutrino mass matrix $M_{\nu}$ is achieved as,
\begin{equation}
\label{eq:2}
\text{Diag}(m_{1},m_{2},m_{3})=U_{PMNS} \ M_{\nu}\ U_{PMNS}^{T},
\end{equation}
where $m_i$(for $i=1,2,3)$ stands for three active neutrino masses. \\
The leptonic mixing matrix is parameterized as,
\begin{equation}
U_{PMNS}={\begin{pmatrix}
	c_{12}c_{13}&s_{12}c_{13}&s_{13}e^{-i\delta}\\
	-s_{12}c_{23}-c_{12}s_{23}s_{13}e^{i\delta}&c_{12}c_{23}-s_{12}s_{23}s_{13}e^{i\delta}&s_{23}c_{13}\\
	s_{12}s_{23}-c_{12}c_{23}s_{13}e^{i\delta}&-c_{12}s_{23}-s_{12}c_{23}s_{13}e^{i\delta}&c_{23}c_{13}\\
	\end{pmatrix}}.P.
\end{equation}
We use abbreviations as $c_{ij}=\cos\theta_{ij}$ , $s_{ij}=\sin\theta_{ij}$ where $\theta_{ij}$ stands for leptonic mixing angles with $i,j=1,2,3 (i\ne j)$. $P$ would be a unit matrix \textbf{1} in the Dirac case but in the Majorana case $P=\text{diag}(1,e^{i\alpha},e^{i(\beta+\delta)})$. $\delta$ and ($\alpha , \beta$) are the Dirac and Majorana CP phases respectively. \\
The inclusion of one extra generation of neutrino along with the active neutrinos lead us to the final $4\times4$ neutrino mixing matrix for the active-sterile mixing as,
\begin{equation}
V\simeq 
\begin{pmatrix}
(1-\frac{1}{2}WW^{\dagger})U_{PMNS} & W \\ -W^{\dagger}U_{PMNS} & 1-\frac{1}{2}W^{\dagger}W
\end{pmatrix},
\end{equation}
where $W=M_{D}M_{R}^{-1}M_{S}^{T}(M_{S}M_{R}^{-1}M_{S}^{T})^{-1}$ is a $3\times1$ matrix guided by the strength of the active-sterile mixing {\it i.e.}, the ratio of $\frac{\mathcal{O}(M_D)}{\mathcal{O}(M_S)}$. In the view of new physics contribution, the trivial $3\times3$ unitary leptonic mixing matrix, $U_{PMNS}$ may slightly deviate from it's generic unitarity behaviour~\cite{Antusch:2006vwa,Akhmedov:2013hec}. Generally the active and sterile mixing lead to non-unitarity in the $U_{PMNS}$ matrix. However, a minimal mixing between the active-sterile neutrinos is considered in our study, which doesn't bother the active neutrino scenario. Moreover, the $U_{PMNS}$ is constrained to be unitary at the $\mathcal{O}(10^{-2})$ level by the current electroweak precision measurements and neutrino oscillation data \cite{Xing:2019vks}.
The sterile neutrino with mass of the order $keV$, can be added to the standard 3-neutrino mass states in NH: $m_1\ll m_2<m_3\ll m_4$ as well as IH: $m_3\ll m_1<m_2\ll m_4$. The diagonalized structure for neutrino mass matrix are modified as $m_{\nu}^{NH}=\text {diag}(0, \sqrt{\Delta m_{21}^{2}}, \sqrt{\Delta m_{21}^{2}+\Delta m_{32}^{2}},\sqrt{\Delta m_{41}^{2}})$ and  $m_{\nu}^{IH}=\text{diag}(\sqrt{\Delta m_{31}^{2}},\sqrt{\Delta m_{21}^{2}+\Delta m_{31}^{2}},0,\sqrt{\Delta m_{43}^{2}})$ respectively for NH and IH mass pattern. Within the generic MES framework, the lightest neutrino mass is zero in both the mass ordering~\cite{Barry:2011wb}. Here, $\Delta m_{41}^{2}(\Delta m_{43}^{2})$ is the active-sterile mass square difference for NH and IH respectively.\\
\begin{table}[h]
	\tbl{The active and sterile neutrino mass matrices and corresponding Dirac ($M_D$), Majorana( $M_R$) and sterile( $M_S$) mass matrices for NH and IH mode. The active-sterile mixing matrices ($W$) and sterile mass for NH and IH mass pattern are also shown in respective columns.}
	{\begin{tabular}{@{}cccc@{}} \toprule
		Structures&  $-m_{\nu}$&\hspace{0.3cm}$m_s$ ($keV$)&$W$\\
		\hline
$\begin{aligned}& ~~~~~~~~~\text{Normal Heirarchy}\\&
	M_R=\begin{pmatrix}
	R_1&0&0\\
	0&R_2&0\\
	0&0&R_3\\
	\end{pmatrix}\\
	& M_{D}= \begin{pmatrix}
	D_1&D_1&D_2+P\\
	0&D_1+P&D_2\\
	P&D_1&D_2\\
	\end{pmatrix}\\
	& M_{S}= \begin{pmatrix}
	G&0&0\\
	\end{pmatrix}\\
	\end{aligned}$& $\begin{pmatrix}
	\frac {D_1^2} {R_2} + \frac {(D_2 + P)^2} {R_3} &\frac {D_1 (D_1 + 
		P)} {R_2} + \frac {D_2 (D_2 + 
		P)} {R_3} &\frac {D_1^2} {R_2} + \frac {D_2 (D_2 + P)} {R_3} \\
	\frac {D_1 (D_1 + P)} {R_2} + \frac {D_2 (D_2 + P)} {R_3} &\frac {(D_1 + 
		P)^2} {R_2} + \frac {D_2^2} {R_3} &\frac {D_1 (D_1 + 
		P)} {R_2} + \frac {D_2^2} {R_3} \\
	\frac {D_1^2} {R_2} + \frac {D_2 (D_2 + P)} {R_3} &\frac {D_1 (D_1 + 
		P)} {R_2} + \frac {D_2^2} {R_3} &\frac {D_1^2} {R_2} + \frac {D_2^2}
	{R_3} \\
	\end{pmatrix}$ &$\simeq\frac{G^2}{\lambda_{1}v}$&${\begin{pmatrix}
	  	\frac{D_1}{G}\\ 0\\ \frac{P}{G}\\
	  	\end{pmatrix}}$	\\
	  \hline
	 $\begin{aligned}
	  &~~~~~~~~~\text{Inverted Hierarchy}\\&
	  M_R=\begin{pmatrix}
	  R_1&0&0\\
	  0&R_2&0\\
	  0&0&R_3\\
	  \end{pmatrix}\\
	  & M_{D}= \begin{pmatrix}
	  D_1&-D_1&D_2+P\\
	  0&-D_1+P&D_2\\
	  P&2D_1&D_2\\
	  \end{pmatrix}\\
	  & M_{S}= \begin{pmatrix}
	  G&0&0\\
	  \end{pmatrix}\\
	  \end{aligned}$
& $ \begin{pmatrix}
\frac {D_1^2} {R_2} + \frac {(D_2 + P)^2} {R_3} &  \frac {D_1(D_1 - 
	P)} {R_2} + \frac {D_2 (D_2 + 
	P)} {R_3} & \frac {-2 D_1^2} {R_2} + \frac {D_2 (D_2 + P)} {R_3} \\
\frac {D_1(D_1 - P)} {R_2} + \frac {D_2 (D_2 + P)} {R_3} &\frac {(D_1 - 
	P)^2} {R_2} + \frac {D_2^2} {R_3} &\frac {-2 D_1 (D_1 - 
	P)} {R_2} + \frac {D_2^2} {R_3} \\
-\frac {2 D_1^2} {R_2} + \frac {D_2 (D_2 + P)} {R_3} & - \frac {2 D_1 (D_1 - 
	P)} {R_2} + \frac {D_2^2} {R_3} &\frac {4 D_1^2} {R_2} + \frac {D_2^2}
{R_3} \\
\end{pmatrix}$	&$\simeq\frac{G^2}{\lambda_{1}v}$
	  &${\begin{pmatrix}
			\frac{D_1}{G}\\ 0\\\frac{P}{G}\\
			\end{pmatrix}}$\\
		\hline
	\end{tabular}\label{tab:nh}}
\end{table}
We have assigned fixed non-degenerate values for the right-handed neutrino mass parameters as $R_1=\times10^{12}$ GeV, $R_2=10^{13}$ GeV and $R_3=5\times10^{13}$ GeV so that they can demonstrate favourable thermal leptogenesis without effecting the neutrino parameters. The mass matrix generated from eq.~\eqref{eq:2} gives rise to complex parameters due to the presence of Dirac and the Majorana phases. As the leptonic CP phases are still unknown, we vary them within their allowed $3\sigma$ ranges (0, 2$\pi$).
We solved the model parameters of the active mass matrix using current global fit $3\sigma$ values for the light neutrino parameters, taken from \cite{Capozzi:2016rtj}.\\
In this work we mainly focus on validating MES to study observable like neutrinoless double beta decay, dark matter and baryogenesis in presence of a $keV$ sterile neutrino ($m_S$) and finally we will try to find correlation among those observable, which we have discussed in following sub-sections. 
\subsection{Neutrino-less Double Beta Decay ($0\nu\beta\beta$):}\label{ndbd}
We assumed that, heavy Majorana neutrinos mediate the observed $0\nu\beta\beta$ process at tree-level. Under the SM framework, the decay amplitude is proportional to \cite{Abada:2018qok, Gautam:2019pce}:
\begin{equation}
\varSigma\ G_f^2\ U_{ei}^2\ \gamma_{\mu}P_R\frac{\cancel{p}+m_i}{p^2-m_i^2}\gamma_{\nu}P_L\simeq \varSigma\ G_f^2\ U_{ei}^2\ \frac{m_i}{p^2}\gamma_{\mu}P_R\gamma_{\nu},
\end{equation}
where, $G_F$ is the Fermi constant, $m_i$ the physical neutrino mass and $p$ is the neutrino virtual momentum such that $p^2= -(125 \text{MeV})^2$. The effective electron neutrino Majorana mass for the active neutrinos in the $0\nu\beta\beta$ process read as, 
\begin{equation}
m^{3_{\nu}}_{eff}= m_1|U_{e1}|^2+m_2|U_{e2}|^2+m_3|U_{e3}|^2,
\end{equation}
The phase "effective $electron$ neutrino" is used as only electrons were involved in the double decay process.
If the SM is extended by $n_S$ extra sterile fermions, the presence of those extra states will modify the decay amplitude which corrects the effective mass as \cite{Benes:2005hn},
\begin{equation}
m_{eff} = \sum_{i=1}^{3+n_S}U_{ei}^2 \ p^2 \frac{m_i}{p^2-m_i^2},
\end{equation}
where, $U_{ei}$ is the $(3+n_S \times 3+n_S)$ matrix with extra active-sterile mixing elements. As we have considered only one sterile state, hence the effective electron neutrinos mass is modified as \cite{Barry:2011wb},
\begin{equation}
m^{3+1}_{eff}= m^{3_{\nu}}_{eff}+m_4|\theta_{S}|^2,
\end{equation}
where, $|\theta_{S}|$ is obtained from the first element of the $R$ matrix and $m_4$ is constrained within [1-18.5] $keV$ \cite{Abada:2018qok} satisfying both $0\nu\beta\beta$ and DM phenomenology under MES framework simultaneously.\\
Many experimental and theoretical progress were made so far and still counting in order to validate the decay process. Interestingly, till date no solid evidences from experiments confirmed $0\nu\beta\beta$ process. However, next-generation experiments \cite{Obara:2017ndb,Artusa:2014lgv,Hartnell:2012qd,Gomez-Cadenas:2013lta,Barabash:2011aa} are currently running in pursue of more accurate limit on the effective mass which might solve the absolute mass problem. Recent results from various experiments give strong bounds on the effective mass $m_{eff}$. Kam-LAND ZEN Collaboration \cite{KamLAND-Zen:2016pfg} and GERDA \cite{Agostini:2018tnm} which uses Xenon-136 and Germanium-76 nuclei respectively gives the most constrained upper  bound upto 90\% CL with $$m_{eff}< 0.06-0.165 \ \text{eV}.$$ Various ongoing and future experiments with their bounds on effective mass are shown in table \ref{teff}. Throughout this work, we consider the future sensitivity of $m_{eff}$ up to 0.01 eV. 
		\begin{table}[h]
			\tbl{Sensitivity of few past and future experiments with half-life in years.}
			{\begin{tabular}{@{}cccc@{}} \toprule
Experiments (Isotope)&$|m_{eff}|$ eV& Half-life (in years)&  Ref.\\
		\hline
		KamLAND-Zen(800 Kg)(Xe-136)&$0.025-0.08$&$1.9\times10^{25}$(90\%CL)&\cite{KamLAND-Zen:2016pfg}\\
		KamLAND2-Zen(1000Kg)(Xe-136)&$<0.02$&$1.07\times10^{26}$ (90\%CL)&\cite{KamLAND-Zen:2016pfg}\\
		GERDA Phase II (Ge-76)& $0.09-0.29$&$4.0\times10^{25}$(90\%CL)&\cite{Agostini:2018tnm}\\
		CUORE (Te-130)& $0.051-0.133$&$1.5\times10^{25}$(90\%CL)&\cite{Artusa:2014lgv}\\
		SNO+ (Te-130)& $0.07-0.14$&$\sim10^{26-27}$&\cite{Hartnell:2012qd}\\
		SuperNEMO (Se-84)&$0.05-0.15$&$5.85\times10^{24}$(90\%CL)&\cite{Barabash:2011aa}\\
		AMoRE-II (M0-100)&$0.017-0.03$&$3\times10^{26}$(90\%CL)&\cite{Bhang:2012gn}\\
		EXO-200(4 Year)(Xe-136)& $0.075-0.2$&$1.8\times10^{25} $(90\%CL)&\cite{Tosi:2014zza}\\
		nEXO(5Yr+5Yr w/Ba Tagging)(Xe-136)& $0.005-0.011$&$\sim10^{28}$&\cite{Licciardi:2017oqg}\\
		\hline
	\end{tabular}\label{teff}}	
\end{table}
\subsection{Dark Matter} \label{sdm}
Since sterile neutrinos cannot thermalize easily, the most straightforward production mechanism is via mixing with the active neutrinos in the primordial plasma \cite{Dodelson:1993je}. 
 Depending upon the production mechanism, one can discard the fact that the mixing of active neutrinos cannot generate sterile neutrinos to behave as a dark matter~\cite{Berlin:2016bdv}. Too large mixing between active-sterile  correspond to a too large DM density however, we can still consider the possibility by considering very small mixing angles\cite{Benso:2019jog, Adhikari:2016bei}.
The DM sterile neutrino production via mixing becomes most efficient at temperatures $T \sim 150-500$ MeV \cite{Dodelson:1993je, Asaka:2006nq, Adhikari:2016bei, Benso:2019jog} resulting in the population of warm DM particles. Resonant production\footnote{Resonantly produced (RP) sterile neutrinos are typically much colder and the dispersion of their momentum distribution is also much smaller than thermal. Therefore, in some sense resonantly produced sterile neutrinos behave as a mixture of a cold and warm DM (CWDM) over some range of scales \cite{Adhikari:2016bei}} results into an efficient conversion of an excess of $\nu_e (\overline{\nu_e})$ into DM neutrinos $S$ \cite{Shi:1998km, Laine:2008pg}. One important thing to keep in mind here is that, the overproduction of dark matter must be avoided to make them experimentally achievable. With proper adjustment of the {\it critical temperature ($T_c$)}\footnote{Temperature at which dark matter production starts.}, we could possibly avoid the overproduction of dark matter abundance. Above the critical temperature, the mixing parameter, $\sin^22\theta_{S}$ from eq. \eqref{dm3} got heavily suppressed, if sterile mass either vanishes or very high at that temperature \cite{Gelmini:2019clw, Benso:2019jog}. If one considers the mixing angle to be a dynamical quantity, then it's not possible to obtain a relic of that quantity. Notwithstanding, in this work, we have considered a tiny static active-sterile mixing angle ($\theta_S<10^{-6}$) such that they remain in the Universe as DM relic~\cite{Yeche:2017upn, Adhikari:2016bei, Baur:2017stq}. The important thing to note here is that, sterile neutrino DM is practically always produced out of thermal equilibrium. Therefore, its primordial momentum distribution is in general, not given by a Fermi-Dirac distribution. Indeed, sterile neutrinos in equilibrium have the same number density as ordinary neutrinos, {\it i.e.}, 112  $cm^{−3} $. With the sterile neutrino mass above $0.4~keV$ would lead to the energy density today $\rho_{sterile, eq}\simeq 45 keV/cm^3$ , which significantly exceeds the critical density of the Universe $\rho_{crit} = 10.5 h^2 keV/cm^3$ . Therefore, sterile neutrino DM cannot be a thermal relic (unless entropy dilution is exploited), and its primordial properties are in general different from such a particle. Detailed discussion on dark matter production mechanism is beyond the scope of this paper, for more comprehensive study one may refer to \cite{Benso:2019jog, Adhikari:2016bei,Gelmini:2019clw}.\\
The most important criterion for a DM candidate is its stability, at least on the cosmological scale. The lightest sterile neutrino is not stable and may decay into SM particles. In the presence of sterile neutrinos, the leptonic weak neutral current is not diagonal in mass eigenstates \cite{PhysRevD.16.1444}, so the $S$ can decay at tree-level via $Z$-exchange, as $S\rightarrow \nu_i\overline{\nu_j}\nu_j$ , where $\nu_i,\nu_j$ are mass eigenstates. The $keV$ sterile neutrino decaying to the SM neutrinos (flavor eigenstates) via $S\rightarrow\nu_{\alpha}\nu_{\beta}\overline{\nu_{\beta}}$ gives the decay width as \cite{PhysRevD.16.1444,PhysRevD.25.766},
\begin{equation}
\Gamma_{S \rightarrow 3_{\nu}}=\frac{G_F^2 m_{S}^5}{96\pi^3}\sin^2 \theta_{S}=\frac{1}{4.7\times 10^{10} sec}\Big(\frac{m_{S}}{50\ keV}\Big)^5\sin^2\theta_{S},
\end{equation}
where, $\theta_{S}$ and $m_{S}$ represents the active-sterile mixing angle and sterile mass respectively. This decay width must give a lifetime of the particle much longer than the age of the Universe. This put a bound on the mixing angle such that, 
\begin{equation}
\theta_{S}< 1.1\times 10^{-7}\Big(\frac{50\ keV}{m_{S}}\Big)^5.
\end{equation}
\begin{figure}
	\includegraphics[scale=0.35]{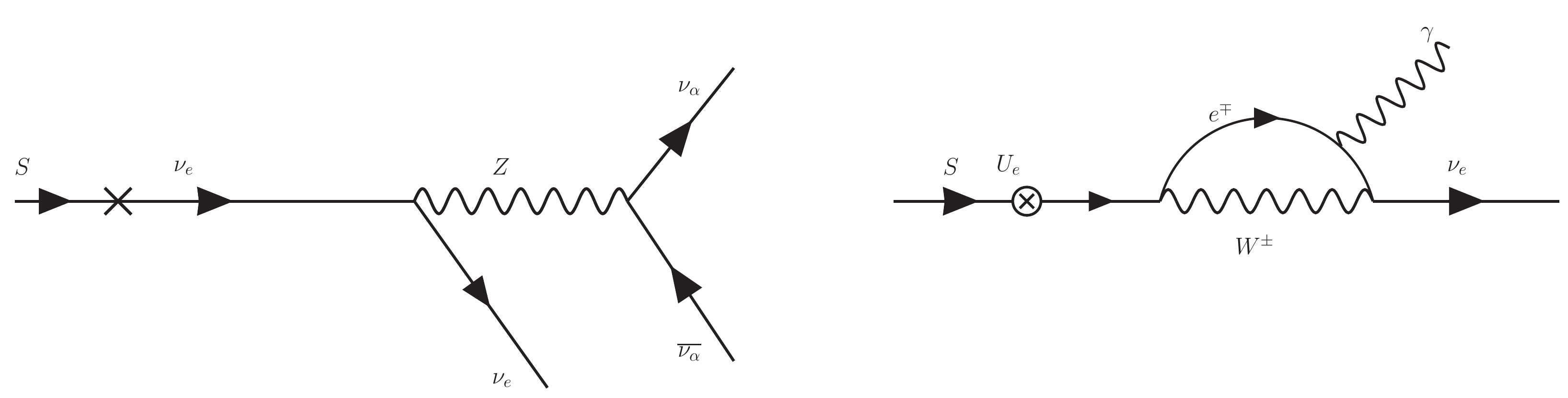}
	\caption{$S\rightarrow\nu_{\alpha}\nu_{\beta}\overline{\nu_{\beta}}\ $ (left) and $S\rightarrow \nu+\gamma$ (right) decay processed of the sterile neutrino \cite{Adhikari:2016bei}. Left figure gives dominant decay channel to three active neutrinos/anti-neutrinos and right figure shows loop mediated radiative decay channel that allows to look for the signal of sterile neutrino DM in the spectra of DM dominated objects.}
\end{figure}
The mass squared difference emerging out of this bound is already much smaller than the current solar mass squared difference. To overcome this short come, either we use another sterile neutrino into the picture or consider one loop-mediated radiative decay process of $S\rightarrow \nu+\gamma$. This would put a stronger bound than the earlier $S \rightarrow 3_{\nu}$ decay process leading to a monochromatic X-ray line signal. However, as discussed in literature \cite{Adhikari:2016bei}, the decay rate is negligible on the cosmological scale because of the small mixing angle. The decay rate for the $S\rightarrow \nu+\gamma$ process is given as \cite{Abada:2014zra,Ng:2019gch}
\begin{equation}
\Gamma_{S\rightarrow \nu\gamma}\simeq 1.32\times10^{-32}\Big(\frac{\sin^2 2\theta_{S}}{10^{-10}}\Big)\Big(\frac{m_{S}}{keV}\Big)^5\label{dm3}
\end{equation}
Relic abundance of the Universe can be worked out starting from the Boltzmann equation. We used results from \cite{Adhikari:2016bei,Abada:2014zra,Ng:2019gch} to check whether our model can verify the observed relic abundance of DM if we consider the sterile neutrino mass in $keV$ range. The working formula for relic abundance is given by,
\begin{equation}
\Omega_{DM}h^2 \simeq 0.3 \Big(\frac{\sin^2 2\theta_{S\nu}}{10^{-10}}\Big)\Big(\frac{m_{S}}{100 keV}\Big)^2,
\end{equation}
with $\theta_{S\nu}$ is the sum of all the active-sterile mixing angles and $m_{S}$ represents the $keV$ ranged sterile neutrino mass. As seen from the above equations, decay rate and the relic abundance depend on the mixing and mass of the DM candidate. Hence, the same set of model parameters that are supposed to produce correct neutrino phenomenology can also be used to evaluate relic abundance and decay rate of the sterile neutrino. 

\subsection{Baryogenesis via Leptogenesis:}\label{lep}
In the early Universe, there is a rapid violation of $B+L$,  at temperatures above the electroweak phase transition (EWPT) \cite{Linde:1977mm}, which converts the lepton asymmetry to baryon asymmetry.  Both the baryon number ($B$) and lepton number ($L$) are independently conserved in the renormalizable SM Lagrangian. However, there are non-perturbative gauge field configurations \cite{Callan:1976je} due to chiral anomaly, which ignites the anomalous $B+L$ violation\footnote{$B-L$ is already conserved.}. This conversion of lepton asymmetry to baryon asymmetry via $B+L$ violation is popularly termed as "sphalerons" \cite{Klinkhamer:1984di} process. In this work, we consider the decay of lightest RH neutrino $\nu_{R1}(L=1)$ to a SM lepton $l(L=1)$ and Higgs $H(L=0)$ and this decay process of $\nu_{R1}\rightarrow l H$, will violate lepton number by two units ($\Delta L=2$). Following the parametrization from~\cite{Davidson:2008bu}, the working formula of baryon asymmetry produced is given by,
\begin{equation}\label{yb}
Y_B=ck\frac{\epsilon_{11}}{g_{*}},
\end{equation}
where, $c$ is the conversion factor. It measures the fraction of lepton asymmetry being converted to baryon asymmetry which is approximately $12/37$.
The term $k$ is the dilution factor due to wash out processes, and this is parametrized as,
	\begin{equation}\label{sk}
	\begin{split}
	k &\simeq \sqrt{0.1K}exp\Big[\frac{-4}{3(0.1K)^{0.25}}\Big], \quad \text{for} \quad K\geq10^6,\\
	&\simeq\frac{0.3}{K(lnK)^{0.6}}, \quad \text{for}\quad 10\leq K\leq 10^6,\\
	&\simeq\frac{1}{2\sqrt{K^2+9}}, \quad\text{for} \quad0\leq K \leq 10.\\
	\end{split}
	\end{equation}
	Here, $K$ is defined as, 
	\begin{equation}
	K=\frac{\Gamma_1}{H(T=M_{\nu_{R1}})}=\frac{(h^{\dagger}h)_{11}M_{\nu_{R1}}}{8\pi}\frac{M_{Planck}}{1.66\sqrt{g_{*}}M^2_{\nu_{R1}}},
	\end{equation}
	where, $\Gamma_1$ is the decay width of $\nu_{R1}$, defined as,  $\Gamma_1=\frac{(h^{\dagger}h)_{11}M_{\nu_{R1}}}{8\pi} $ and the Hubble constant at $T=M_{\nu_{R1}}$ is defined as $H(T=M_{\nu_{R1}})=\frac{M_{Planck}}{1.66\sqrt{g_{*}}M^2_{\nu_{R1}}}$. 	The quantity $g_{*}$ is the mass-less relativistic degree of freedom in the thermal bath and it's value is approximately around $110$.	The most important term is the lepton asymmetry term $"\epsilon_{11}"$ that produced by the decay of the lightest RH neutrino $\nu_{R1}$. \\Decay of the $\nu_{R1}$ must have a lepton number violating process with different decay rate to final state with particle and anti-particle, otherwise the lepton asymmetry would be vanished. Asymmetry produced by the decay of $\nu_{R1}$ in lepton flavor $\alpha$ produced is defined as,
	\begin{equation}
	\epsilon_{\alpha\alpha}=\frac{\Gamma(\nu_{R1} \rightarrow l_{\alpha}H)-\Gamma(\nu_{R1} \rightarrow \overline{l}_{\alpha}\overline{H})}{\Gamma(\nu_{R1} \rightarrow l H)+\Gamma(\nu_{R1} \rightarrow \overline{l}\overline{H})},
	\end{equation}
	where $\overline{l}_{({\alpha})}$ is the antiparticle of $l_{({\alpha})}$ and $H$ is the Higgs doublet. With non-degenerate RH mass\footnote{For degenerate RH mass with mass spiting equal to decay width lead to the case of  resonant leptogenesis.}, we carried out our numerical analysis from the work of \cite{Joshipura:2001ya} and obtained the asymmetry term as,
	\begin{equation}
	\begin{split}
	\epsilon_{\alpha\alpha}= & \frac{1}{8\pi}\frac{1}{[h^{\dagger}h]_{11}}\sum_{j}^{2,3}\text{Im} {(h_{\alpha 1}^{*})(h^{\dagger}h)_{1j}h_{\alpha j}} g(x_{j})\\
	& + \frac{1}{8\pi}\frac{1}{[h^{\dagger}h]_{11}}\sum_{j}^{2,3}\text{Im} {(h_{\alpha 1}^{*})(h^{\dagger}h)_{1j}h_{\alpha j}} \frac{1}{1-x_j},\\
	\end{split}\label{epsi}
	\end{equation} 
	where $x_j\equiv \frac{M_j^2}{M_1^2}$ and within the SM $g(x_j)$ is defined as ,
	\begin{equation}
	g(x_j)=\sqrt{x_j}\Big(\frac{2-x_j-(1-x_j^2)\text{ln}(1+x_j/x_j)}{1-x_j}\Big).
	\end{equation}
	 When we take sum over $\alpha$, the second line from equation \eqref{epsi} violates single lepton flavors, however, it conserves the total lepton number and it vanishes.
	\begin{equation}\label{ep}
	\epsilon_{11}\equiv\sum_{\alpha}\epsilon_{\alpha\alpha}=\frac{1}{8\pi}\frac{1}{[h^{\dagger}h]_{11}}\sum_{j}^{2,3}\text{Im}{[(h^{\dagger}h)_{1j}]^2} g(x_{j})
	\end{equation}
	The $h$ used here is the Yukawa matrix generated from the Dirac neutrino mass matrix and the corresponding index in the suffix conveys the position of respective matrix element.
	
We constructed the Yukawa matrix from the solved model parameters $D_1,D_2$ and $P$ , which is related to the $3\times3$ Dirac neutrino mass matrix. The $K$ value within our study fell in the range $10\leq K\leq 10^6$; thus, we have to go for the second parametrization of the dilution factor from equation \eqref{sk}.Now, the baryon asymmetry of the Universe can be calculated from equation \eqref{yb}, followed by the evaluation of lepton asymmetry using the equation \eqref{ep}.
\section{Results} \label{result}
\begin{figure}\centering
	\includegraphics[scale=0.45]{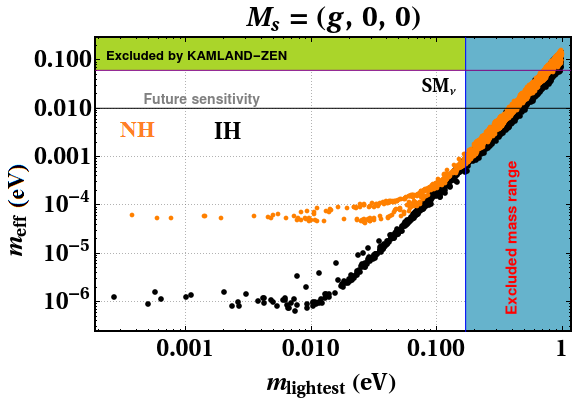}\\
	\includegraphics[scale=0.4]{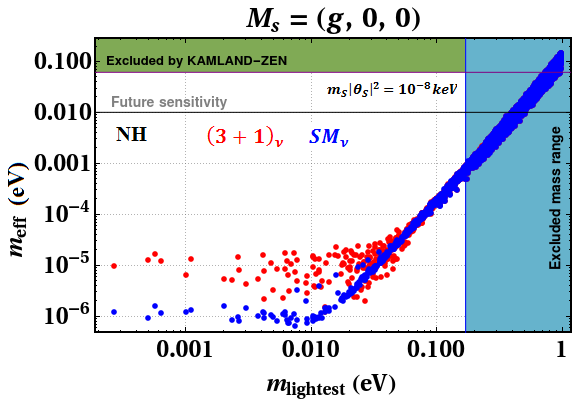}
	\includegraphics[scale=0.4]{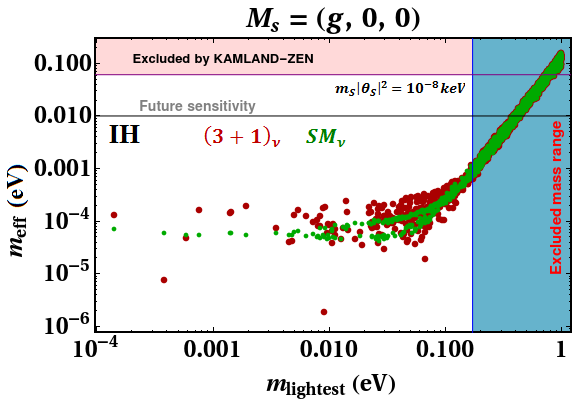}
	\caption{Variation of effective neutrino mass vs. the lightest neutrino mass. The upper plot represents the contribution from the active neutrinos only. The lower two plots represent NH and IH respectively, for both active and active+sterile contributions. Horizontal gray line and vertical blue line represent the future $m_{eff}$ bound and sum of the active masses respectively. In presence of sterile neutrino, a much wider and significant impact is visible on the bottom plots. In both the mass orderings we have fixed the mixing element $m_S|\theta_{S}|^2=10^{-8}~keV$.  }\label{eff}
\end{figure}
\begin{figure}
	\includegraphics[scale=0.3]{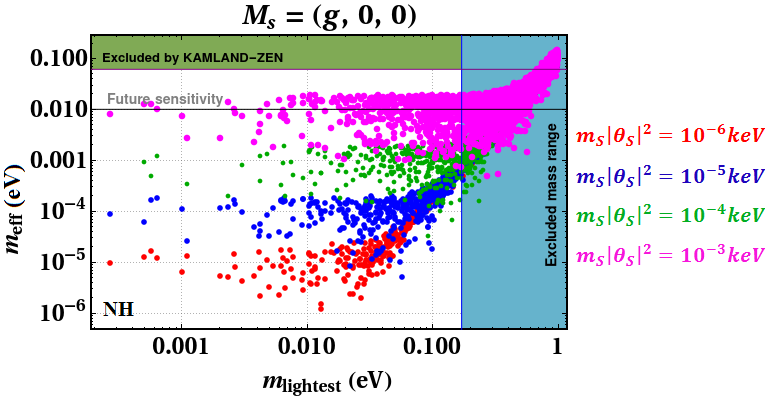}
	\includegraphics[scale=0.3]{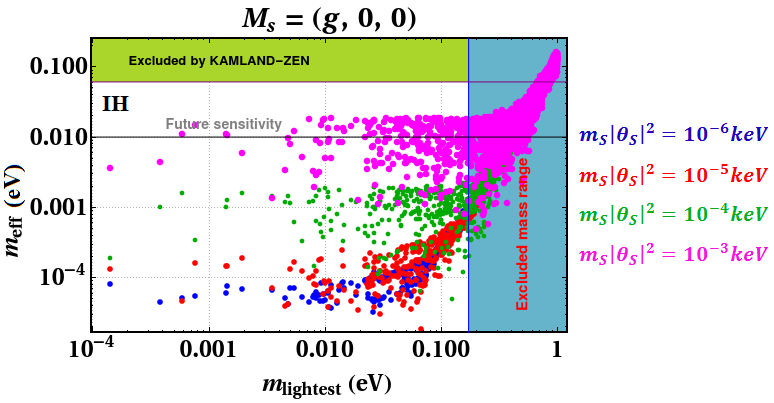}
	\caption{Variation of effective mass for different ranges of active-sterile mixing angle. The left plot represents the NH mode while the right one represents the IH mode. In both the mass patterns, $m_S|\theta_{S}|^2>10^{-4}~keV$ fails to satisfy the future sensitivity bound of effective mass. This result gives upper bound on the active-sterile mixing element $"|\theta_{S}|^2"$ from the $0\nu\beta\beta$ study. }\label{u4}
\end{figure}
Under the hypothesis that future experiments will verify the existence of at least one heavy sterile neutrino in $keV$ range, we work out the possibility of its effect on $0\nu\beta\beta$ and verifying the fact that this sterile neutrino could behave as DM within the mass range of (1-18.5) $keV$. We have plotted effective neutrino mass ($m_{eff}$) against the lightest neutrino mass ($m_{lightest}$) in fig. \ref{eff}. The horizontal gray line gives the future sensitivity of upper bound on effective mass up to $10^{-2}$ eV, and the vertical blue line gives the upper bound on the sum of the active neutrino masses ($0.17$ eV). In the upper part of fig. \ref{eff}, NH (black) and IH (orange) contributions are coming only from the active neutrinos, whereas, in lower two figures, NH and IH contributions are shown separately in the presence of $m_S$. In presence of the sterile neutrino, one can observe a wider and improved data range in both the mass ordering. These extra contributions and improvements in effective mass are due to the sterile neutrino mass ($m_S$) and the active-sterile mixing ($\theta_{S}$). In fig. \ref{u4}, we completed the same analysis of $m_{eff}$ vs. $m_{lightest}$ for different orders of active-sterile mixing element. Very interesting results are observed from both NH and IH mode. For $m_4|\theta_{S}|^2>10^{-4}$ $keV$, $0\nu\beta\beta$ fails the future experimental bound. From these results, we get the upper bound on the active-sterile mixing angles and it is also obvious from the fact that the active-sterile mixing element must be very small otherwise there would be an overproduction of dark matter in our Universe \cite{Benso:2019jog}.\\
\begin{figure}\centering
	\includegraphics[scale=0.3]{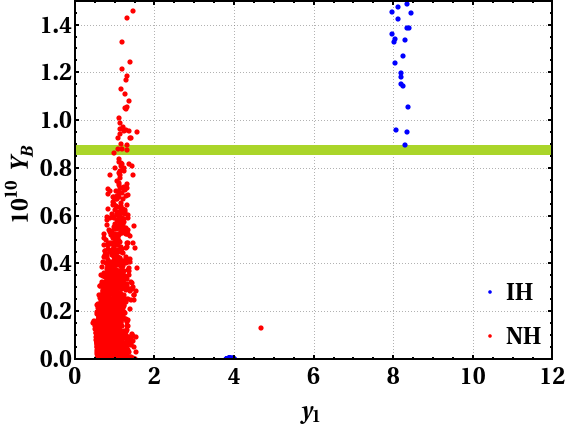}
	\includegraphics[scale=0.3]{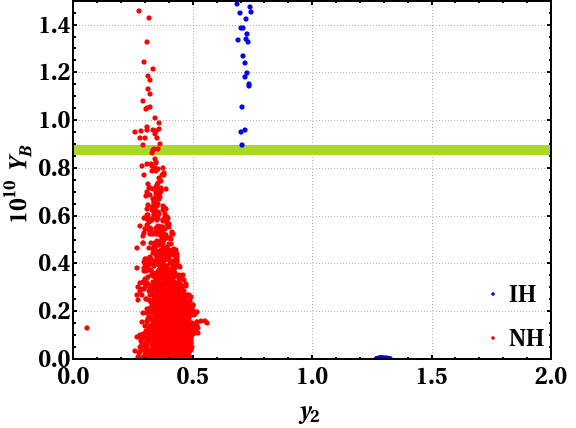}
	\includegraphics[scale=0.3]{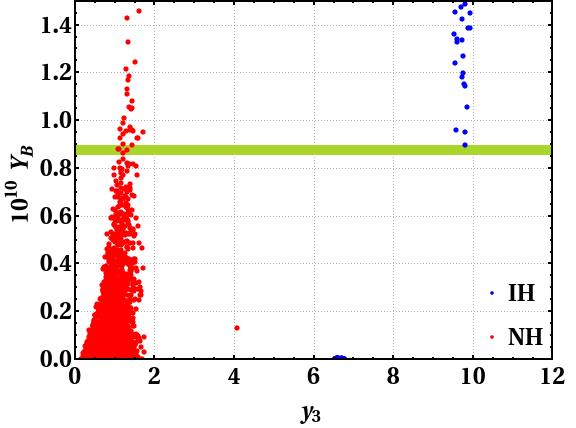}
	\caption{Variation of Yukawa coupling with BAU in both the mass ordering. Solid green band represents the current BAU value within 3$\sigma$ range, which is $(8.7\pm0.06)\times 10^{-11}$. Red and blue points represents NH and IH mass ordering respectively. Yukawa couplings ($y_1,y_2,y_3$) satisfying BAU results are achieved around $\mathcal{O}(10^{-1}-1)$, which are also in accessible range for dimension-5 frameworks. Stringent regions on the Yukawa couplings are due to recent bounds on the light neutrino parameters. }\label{yib}
\end{figure}
Variation of Yukawa couplings with the baryogenesis result are shown in fig. \ref{yib}. The red dots represent NH and the blue dots represent IH respectively. The green bar gives the current allowed 3$\sigma$ value of BAU. As BAU value is highly sensitive to the experimental results, very narrow regions are observed in both the mass ordering satisfy baryogenesis in our model and NH shows more favorable results when we vary BAU with the Yukawa couplings. In current dimension-5 scenario, the Yukawa coupling of  $\mathcal{O}(10^{-2}-1)$~\cite{Abada:2007ux, Ibarra:2010xw, Das:2017nvm} are in acceptable range. Strong constrained regions in fig. \ref{yib} are due to the bounds on light neutrino parameters imposed by the Yukawa matrix involved in the baryogenesis calculation. These constrained regions of Yukawa couplings also put stringent bounds on the light neutrino parameters. For example, within NH, for large $y_3(\ge2.0)$, $\Delta m_{31}^2$ value exceed the current upper bound of $3\sigma$ value, whereas small $y_3(\le 0.2)$, $\Delta m_{31}^2$ value goes beneath the lower $3\sigma$ bound.
\begin{figure}
	\begin{minipage}[t]{0.47\textwidth}
		\includegraphics[scale=0.28]{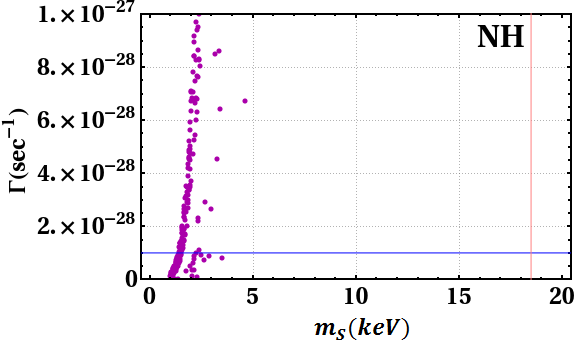}
		\includegraphics[scale=0.28]{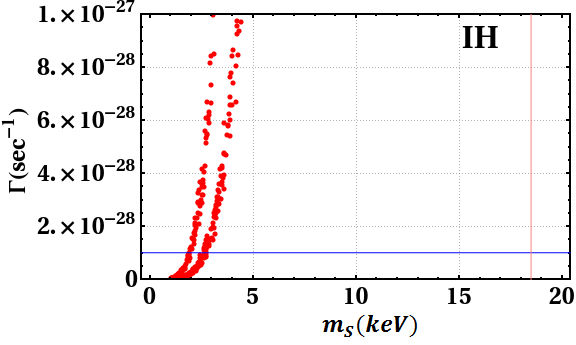}
	\end{minipage}
	\begin{minipage}[t]{0.37\textwidth}
		\includegraphics[scale=0.25]{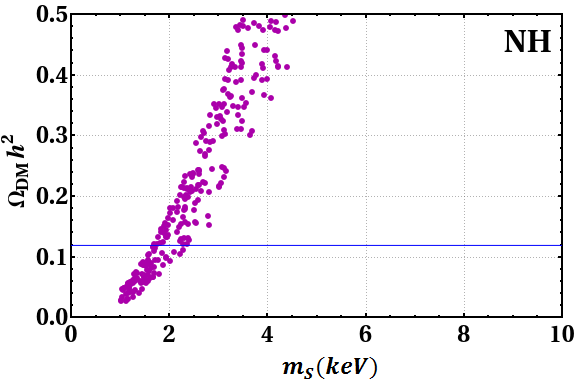}
		\includegraphics[scale=0.25]{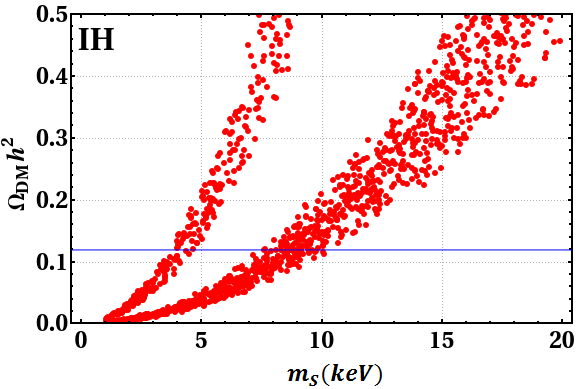}
	\end{minipage}
	\caption{Variation of decay width ($\Gamma$) and the relic abundance ($\Omega_{DM}h^2$) of the Universe vs. the sterile neutrino mass ($m_S$). Sterile neutrino to behave as a dark matter, $m_S\simeq7.1~keV$ and lifetime $\tau_{DM}(\Gamma^{-1})\simeq10^{27.8\pm0.3}~sec$ are suggested from the recent discovery of $E=3.55~keV$ line in the X-ray spectra of galaxy cluster\cite{Bulbul_2014,Boyarsky:2014ska}. For decay width plots, the blue lines give the upper limit of the decay width which is considered to be $\Gamma<10^{-28} sec$. On the other hand for relic abundance plots, the solid blue line gives the current best fit value for relic abundance of a particle to behave as a dark matter ($\Omega_{DM}h^2=0.119$). $m_S$ around 1-3 $keV$ is consistent with NH mode and for IH mode the mass ranges are different while satisfying both the decay width and relic abundance. In IH mode, $m_S$ around 1-3 $keV$ is consistent with the upper bound of decay width, while $m_S$ between 4-10 $keV$ satisfy the current relic abundance value.}\label{dm}
\end{figure}
Parallel to the $0\nu\beta\beta$ study, we have also examined dark matter signature of the $keV$ sterile neutrino in fig. \ref{dm}. Decay width ($\Gamma$) and relic abundance of the sterile neutrino ($\Omega_{DM}h^2$) are plotted against the sterile mass ($m_{S}$) for both the mass ordering. Sterile neutrinos to behave as a DM, their lifetime must be greater than the age of the Universe so that their remnants remain in the Universe; hence, the decay width of the particle must be very less. In our study, we have considered the upper limit of decay width to be less than $10^{-28}\ (sec^{-1})$. The sterile neutrino mass is considered in a narrow region, i.e., $(1-18.5)$ $keV$ to be a relic particle. Relic abundance obtained in both the mass ordering, satisfy the proper bound with different $m_S$ ranges. The allowed mass range for the sterile neutrino is very narrow $(1-3)$ $keV$ in case of NH mode, while a broad mass spectrum satisfies the upper relic abundance bound in IH mode $(1-10)$ $keV$. 
Recent results suggested $m_S\simeq7.1~keV$ and lifetime $\tau_{DM}\simeq10^{27.8\pm0.3}$ sec. \cite{Boyarsky:2014ska}.
Even though the decay width and the relic abundance of the sterile neutrino are satisfied in both the mass ordering, NH results are more consistent with sterile mass within (1-3) $keV$. On the other hand in case of IH mode, relic abundance limit is within the sterile mass range from 4 $keV$ to 10 $keV$, while decay width is satisfied with a small mass up to 3 $keV$.\\
\begin{figure}
	\includegraphics[scale=0.3]{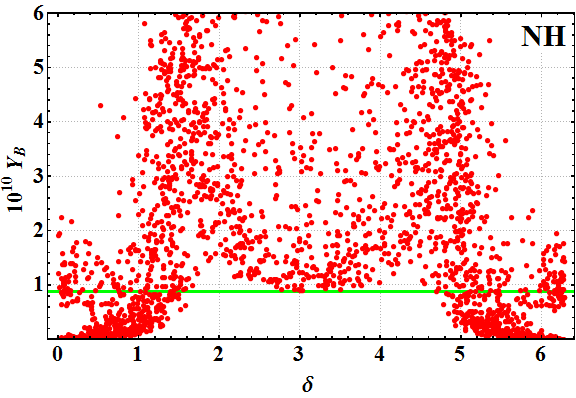}
	\includegraphics[scale=0.3]{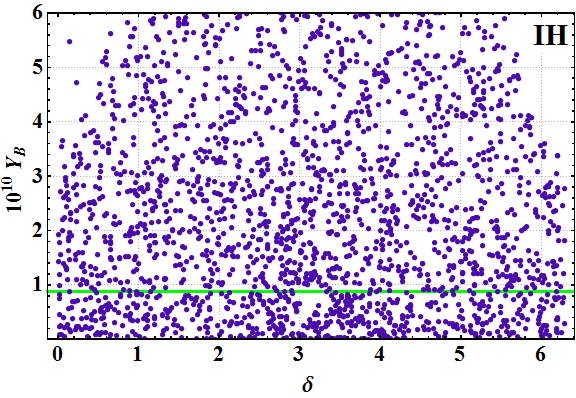}
	\caption{Variation of the Dirac delta phase with $Y_B$ in both the mass ordering. The solid green band represents the current BAU value, $Y_B=(8.7\pm0.06)\times 10^{-11}$. Both the mass orderings satisfy baryogenesis in our model and correlate with $\delta$ }\label{lep1}
\end{figure}
\begin{figure}
	\includegraphics[scale=0.23]{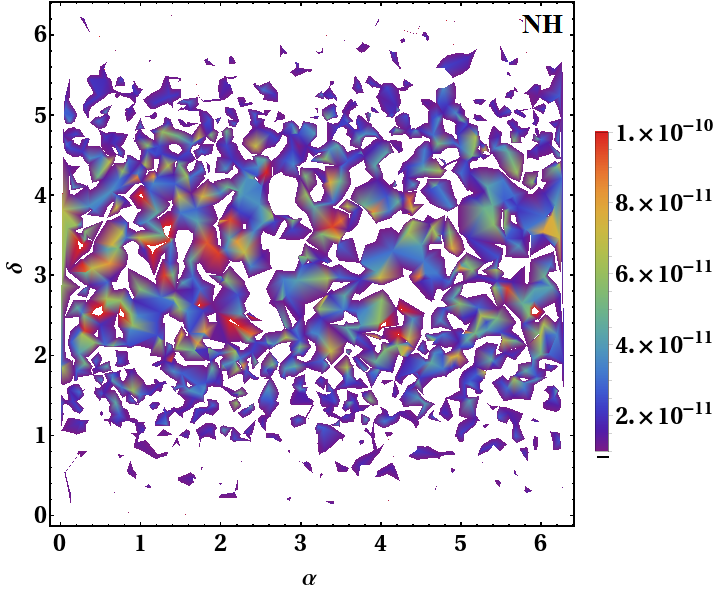}
	\includegraphics[scale=0.23]{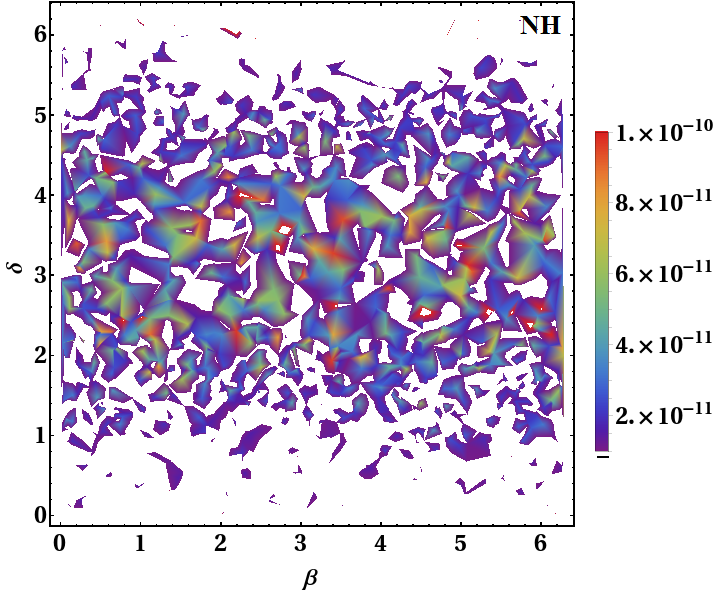}
	\caption{Projection of BAU value ($Y_B$) between Dirac CP-phase ($\delta$) along X-axis and Majorana phases ($\alpha$ and $\beta$ respectively) along Y-axis. Current BAU value range is around the red-orange colour band and this constrains the Dirac CP phase ($\delta$) in between the numerical values (2.0-4.0).}\label{delta}
\end{figure}
\begin{figure}
	\includegraphics[scale=0.3]{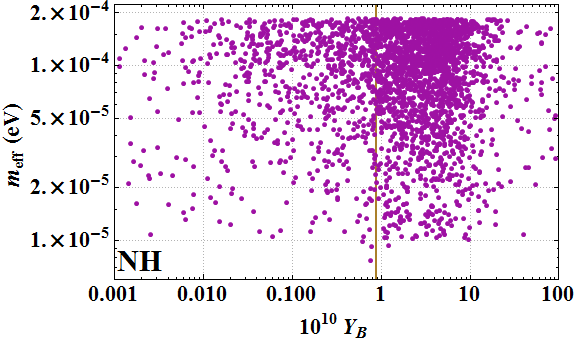}
	\includegraphics[scale=0.3]{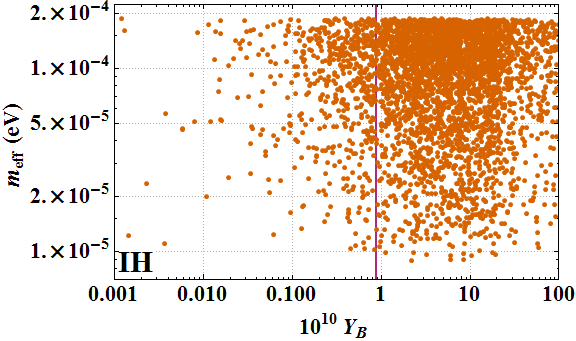}
	\caption{Correlation between effective neutrino mass ($m_{eff}$) with $Y_B$ in NH and IH respectively. Solid vertical band represents the current BAU value, which is $(8.7\pm0.06)\times10^{-11}$. In both the cases, $m_{eff}$ lie well below the current upper bound and the solid vertical line indicates successful execution of baryogenesis and $0\nu\beta\beta$ in the model.}\label{ebau}
\end{figure}
\begin{figure}
	\includegraphics[scale=0.24]{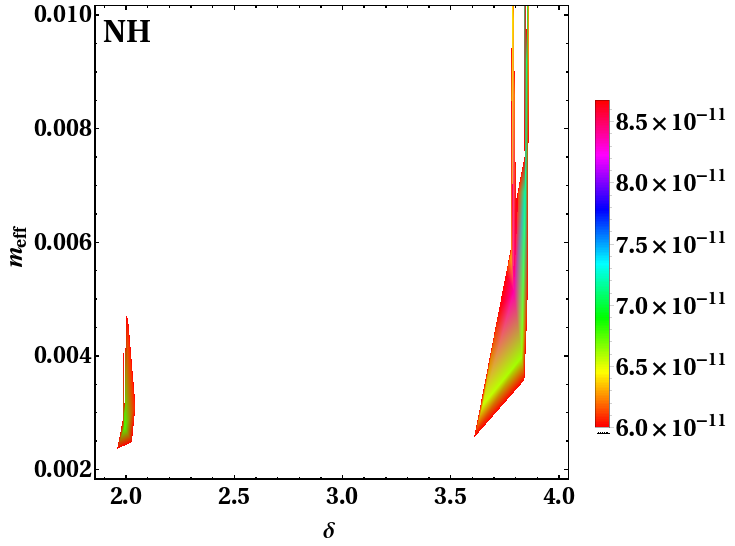}
	\includegraphics[scale=0.24]{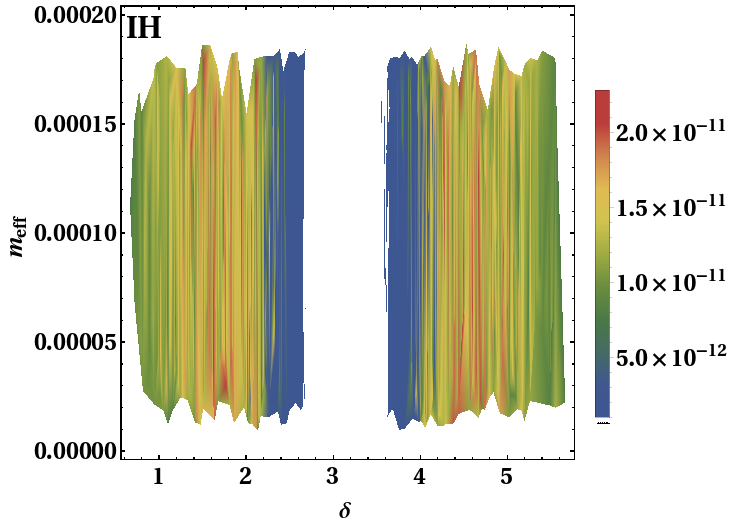}
	\caption{Projection of BAU value ($Y_B$) in a frame representing effective mass ($m_{eff}$) along Y-axis and the Dirac CP-phase ($\delta$) along X-axis. A precise constrained range for the Dirac CP-phase value around 3.5-4.0 is obtained for NH mode. Whereas IH mode failed to reflect the exact $Y_B$ value. }\label{deb}
\end{figure}
\begin{figure}
	\includegraphics[scale=0.25]{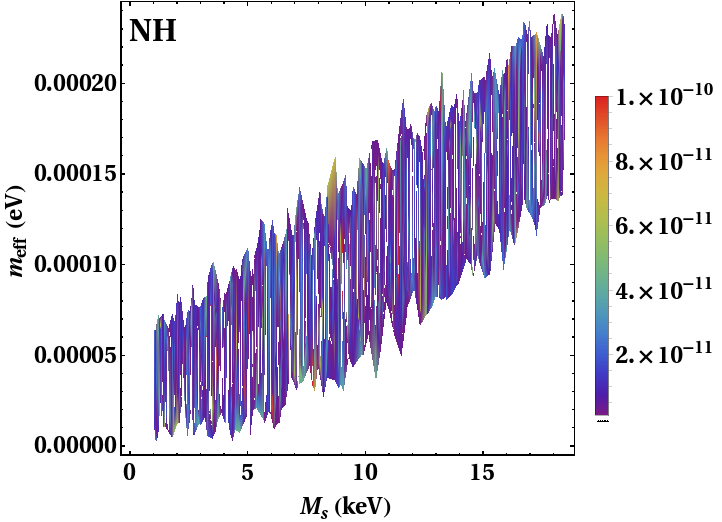}
	\includegraphics[scale=0.24]{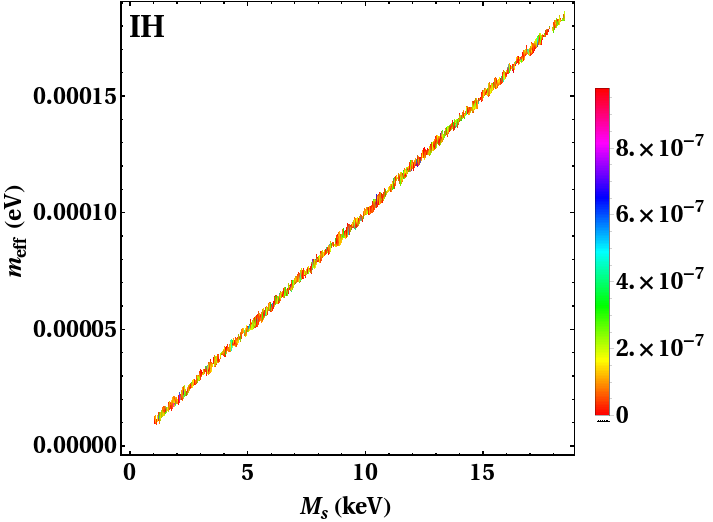}
	\caption{Projection of $Y_B$ in a frame representing effective mass ($m_{eff}$) along Y-axis and sterile mass ($m_S$ in $keV$) along X-axis. Here also $Y_B$ is much higher than its current bound in the IH mode and fails to correlate with $m_{eff}$ and $m_S$. On the other hand NH was able to project the BAU bound successfully, however, very small regions are observed. }\label{seb}
\end{figure}
We also study the results for baryogenesis via the mechanism of thermal leptogenesis and showed a co-relation among other observable. In fig. \ref{lep1}, we varied the Dirac delta phase ($\delta$) with the baryon asymmetry of the Universe calculated in our model in both the mass orderings. Both these results show the validity of BAU within our model. Similar results can be seen from fig. \ref{delta}, where we project BAU in between Dirac and Majorana phases for NH mode only. Co-relation among the effective neutrino mass in the presence of $keV$ sterile neutrino with the BAU is also shown in fig. \ref{ebau}. Since the Dirac CP phase has influence in $0\nu\beta\beta$ as well as in BAU, we added contour plot with $m_{eff}$ and $\delta$ along the axes and projected BAU value in the Z-plane in fig. \ref{deb}.  Constrained regions in the Dirac CP phase are observed in both the mass orderings. As we can see from the legends on right-hand side of the figure, that IH pattern failed to project the current observed value of BAU in a frame of $m_{eff}$ Dirac-CP phase $\delta$. We also present another contour plots in fig. \ref{seb}, where a measured BAU is projected in the frame between sterile neutrino mass ($m_S$) and effective electron neutrino mass. Since BAU results are very sensitive to the experiments, very narrow regions are observed. We can see that IH mode is almost ruled out in presence of sterile mass and mixing while we get some region satisfying current BAU bound in NH mode. 
\section{Conclusion} \label{conc}
In this work, we study the viability of $keV$ sterile neutrino to behave as a warm dark matter and giving an observable effect in $0\nu\beta\beta$ and baryogenesis via the mechanism of thermal leptogenesis. We use $A_4$ based flavor model with discrete $Z_4\times Z_3$ to construct desired Yukawa coupling matrices. Here, the Dirac neutrino mass $M_D$ is a $3\times3$ complex matrix, the Majorana mass matrix $M_R$, which arises due to the coupling of right-handed neutrinos is also a $3\times 3$ complex symmetric diagonal matrix with non-degenerate values. A singlet gauge fermion $S$ is considered which couples with the right-handed neutrino, hence produces a single row  $1\times 3$ $ M_S$ matrix with one non-zero entry. The Dirac neutrino mass matrix, $M_D$, is modified using a matrix, $ M_P$, which is generated via the same fashion as $M_D$ to make the active mass matrix $\mu-\tau$ asymmetric.
 Few interesting points based on the results are discussed as follows,
\begin{itemize}
	\item Presence of an extra heavy sterile flavor has a significant impact on effective neutrino mass. One can find a broader effective mass range in the active-sterile case than the active neutrino case.  Normal hierarchy (NH) is more favourable than the inverted hierarchy (IH) mode for $0\nu\beta\beta$ in this MES framework.
	\item Consequential bound on active-sterile mixing angle is obtained for future sensitivity in effective mass from fig. \ref{u4}, which restricts the upper bound on the mixing element up to $10^{-4}$ for $|\theta_{S}|^2$.
	\item In fig. \ref{yib}, strongly constrained regions for the Yukawa couplings are obtained through baryogenesis calculation, which by the by gives strict bounds on the choice model parameters.
	\item Dark matter analysis results from decay width and relic abundance restricts 
	sterile neutrino mass within few $keV$ to behave as dark matter. Among different bounds for thermal relic mass for the sterile neutrino, very few results
	are consistent with X-ray observations. Lyman-$\alpha$ forest of high resolution quasar spectra with hydrodynamical N-body simulations gives bounds ranging from $m_S \geq 1.8$ $keV$ to $m_S\geq 3.3$ $keV$ \cite{Viel:2006kd,Seljak:2006qw,Viel:2013apy}. Regardless, these bounds may vary depending upon various uncertainties effecting the constraints \cite{Schultz:2014eia}. Within MES framework, NH predicts sterile mass range from ($1-3$ $keV $) and IH results for relic abundance gives mass up to $10$ $keV$ while the decay width constraints the mass within 3 $keV$.  Hence, from these results we come to a conclusion that with current bounds on hand, sterile neutrino as a dark matter in minimal extended seesaw is still an unsettled aspect. A deeper discussion with new bounds on $keV$ sterile neutrino may resolve these issues, which is left for future studies.
	\item BAU is satisfied in this framework, and NH shows more efficient in producing the observed matter-antimatter density than IH pattern. This model also successfully correlate $0\nu\beta\beta$ with BAU result,
	which can be found in fig. \ref{ebau}. Projection of BAU on a plane in between effective mass and Dirac CP phase, $\delta$ gives significant remark in our study. In fig. \ref{deb}, one can find that BAU results are constraining $\delta$ in both the mass ordering and NH results are more favourable with current BAU value than the IH. Within NH, $\delta$ is tightly constrained in between ($2.0-4.0$) value.
	\item Projection of BAU with sterile mass and effective mass in presence of sterile neutrino
	in fig. \ref{seb} gives an unsatisfactory remark while observing IH. Hence, IH fails to correlate them in a single frame. In spite of the fact that, BAU value is very
	small, NH manages to project the value along with $keV$ sterile neutrino. 
	\item  In NH mode within this model, a constrained bound on the Dirac CP-phase is obtained from baryogenesis study, which can be seen in the density plot of fig. \ref{delta} with Majorana phases in the X-axis. Majorana phases cover the whole $0-2\pi$ range, whereas the Dirac CP-phase is constrained between the value ($2.0-4.0$) satisfying observed BAU value.
\end{itemize}
In conclusion, the MES mechanism is analyzed in this work, considering a single flavor of a sterile neutrino in a $keV$ scale. Along with the active and sterile mass generation, this model can also be used to study the connection between effective mass in neutrinoless double beta decay ($0\nu\beta\beta$) in a wider range of sterile neutrino mass, simultaneously addressing the possibility of $keV$ scale sterile neutrino as dark matter particle. Although, results on $keV$ sterile neutrino as a dark matter candidate is still on the verge of uncertainty within the framework of MES. We keep an optimistic hope to get better bounds from future experiments which may establish the same within MES. Results from baryogenesis via the mechanism of thermal leptogenesis are also checked and verified within this model. Finally, we have correlated all these observable under the single framework. Results in NH mass pattern shows better consistency than IH pattern.
\section*{Acknowledgements}
{This work is supported by the Department of Science and Technology,  Government of India under project number EMR/2017/001436.}
\bibliographystyle{ws-ijmpa}
\bibliography{kevdm.bib}
\end{document}